\documentclass[twocolumn,prl,showpacs]{revtex4}
\usepackage{graphicx}
\usepackage{dcolumn}
\usepackage{bm}
\usepackage{amsmath}
\usepackage{hyperref}
\begin{document}

\preprint{}
\title{Monopole Current and  Unconventional Hall Response on Topological Insulator}
\author{Jiadong Zang$^{1,2}$ and Naoto Nagaosa$^{1,3}$}
\affiliation{
$^1$ Department of Applied Physics, University of Tokyo, Tokyo 113-8656, Japan \\
$^2$ Department of Physics, Fudan University, Shanghai 200433, China\\
$^3$ Cross Correlated Materials Research Group (CMRG), ASI, RIKEN,
WAKO 351-0198, Japan }
\date{\today}

\begin{abstract}
We study theoretically the charged current above a topological
insulator (TI) separated by a ferromagnetic insulating layer. An
unconventional Hall response occurs in the conducting layer on top
of the TI which approaches to a constant value independent of $R$
for $R\ll\ell$ and decays with $\propto R^{-1}$ for $R\gg\ell$,
where $R$ is the separation between TI and conducting layer and
$\ell$ is the screening length. In the comoving frame, it can be
interpreted as a monopole current attached to the TI surface. The
same mechanism gives the Hall response and deflection of the
electron beam injected to the surface of insulating ferromagnet. A
realistic estimate of an order of magnitude shows that both effects
give reasonably large signal experimentally accessible.
\end{abstract}

 \maketitle



%

%



Topological insulator (TI) is a new state of matter realized in the
noninteracting electron systems, i.e., the nontrivial band structure
characterized by the "twist" of the Bloch wavefunction in the
momentum
space\cite{Bernevig2006a}\cite{Fu2007a}\cite{Kane2005b}\cite{Hsieh2008a}.
As in the case of quantum Hall system, there is a gap in the bulk
states, and the manifestation of the nontrivial topology appears on
the surface (edge) of the three (two) dimensional TI\cite{Qi2008a}.
In the case of 3D TI, there appears the helical Dirac fermions on
the surface, which is robust against the disorder. This helical
metal state is expected to produce the several novel properties such
as the topological magneto-electric (TME) effect\cite{Qi2008a}, and
an image magnetic monopole when a charge is put above the
TI\cite{Qi2009a}. For these effect to be observed, the time-reversal
symmetry breaking is needed, which can be achieved by the
ferromagnetic thin layer attached on top of TI, which induces the
exchange coupling and the gap to the surface Dirac fermion and its
anomalous Hall effect (AHE). Especially, when the Fermi energy lies
within the gap, the Hall conductance is predicted to be quantized as
$\pm e^2/(2h)$, i.e., half of the conductance unit. When this
condition is satisfied, the distribution of the magnetic field
outside of the TI is that given by the image magnetic monopole
inside the TI. However, in realistic situation, the Fermi energy is
rather difficult to control, and lies within the finite density of
states of the surface Dirac fermions even with the gap opens by the
exchange coupling.

When the TI surface is gapped, and the Fermi surface exactly lies in
the gap, the effective electromagnetic response of a 3D TI can be
described by $\theta$-term in the Lagrangian\cite{Qi2008a},
\begin{equation}
\mathcal{L}_{eff}=\frac{\theta}{2\pi}\frac{\alpha}{4\pi}\epsilon^{\mu\nu\rho%
\sigma}F_{\mu\nu}F_{\rho\sigma}
\end{equation}
where $F_{\mu\nu}$ is the electromagnetic field strength, and
$\alpha$ is the fine structure constant. $\theta=0$ for conventional
insulator, while $\theta=\pm\pi$ for TI. Concerning the chiral
anomaly, the sign above is decided by the direction of a magnetic
field or magnetization on the TI surface. This nonvanishing $\theta$
leads to the topological magneto-electric effect of TI. As a result,
when a pure charge is placed on the top of a chirality fixed TI
surface, its electric field induces a magnetic field. It's amazing
that this magnetic field lines originate from the charge's mirror
position with respect to the TI surface. In this sense, we may say
that a charge would induce a monopole in the mirror with the help of
TI\cite{Qi2009a}. Without losing the generality, assume the unity
dielectric constant and magnetic permeability of the TI. The
monopole strength of the induced monopole in SI units is given by
$g=\frac{2\alpha\mu_0c}{%
(4+\alpha^2)}q=\frac{e^2}{2h}\frac{2\mu_0}{\varepsilon_0(4+\alpha^2)}q$,
with $\alpha$ being the fine structure constant.

\begin{figure}[tbp]
\begin{center}
\scalebox{0.8}{
\includegraphics[width=10cm,clip]{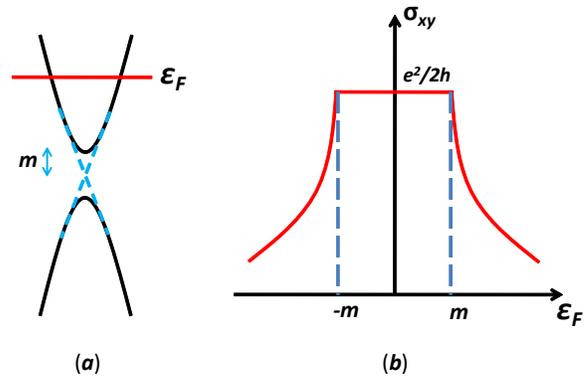}
}
\end{center}
\caption{ Sketch of ({\bf a}) the relative positions of Fermi
surface $\varepsilon_F$ and the magnetic gap $m$ and ($\bf b$)
relation between Hall conductance and the Fermi surface. When the
Fermi surface lies in the magnetic gap $m$, the conductance is
quantized as half the conductance quanta. However, when the Fermi
surface is pushed outside the gap, $\sigma_{xy}$ decays inversely
proportional to $\varepsilon_F$.} \label{fermi}
\end{figure}


Phenomenologically TME can be best interpreted as the quantum Hall
effect on the TI surface by applying the bulk-edge correspondence.
In the presence of a perpendicular magnetic field, quantum Hall
effect with half conductance quanta is realized for the chiral
liquid on TI surface\cite{Qi2008a}, namely,
$\sigma_{xy}=\frac{e^2}{2h}$. Therefore, in-plane component of the
electric field induced by a static charge generates a circulating
Hall current. TME is nothing but the orbital magnetization due to
this Hall current.

However in reality, the Fermi level does not come across the
magnetic gap, but lies within in the finite density of states
(Fig.\ref{fermi}(a)). Experimentally the approachable magnetic gap
is $10$K at most, and it is difficult to push the Fermi surface
inside the gap. On the other hand, it's quite possible that the
Fermi surface lies about $100$K ($10\mathrm{meV}$) above or below
the Dirac point.

Microscopically, the helical liquid on gapped TI surface is given by
\cite{Hsieh2008a}\cite{Qi2008a}\cite{Chen2009a}:
\begin{equation}
H={\bf k}\cdot(\sigma\times\hat{z})+m\sigma_z
\end{equation}
where $m$ is the strength of perpendicular magnetization. The
half-quantized Hall effect holds true only when the Fermi surface
lies in the gap opened by the chirality fixing field. When the Fermi
surface lies away from the gap, the conductance would not be half
quantized anymore. Employing TKNN formula\cite{Thouless1982a}, we
get the Hall conductance
\begin{equation}
\sigma_{xy}=\frac{m}{\varepsilon_F}\frac{e^2}{2h}
\end{equation}
where $\varepsilon_F$ is the Fermi surface, see Fig.\ref{fermi}(a).
It explicitly shows that the transverse conductance is suppressed by
a factor of $m/\varepsilon_F$, so is TME and the monopole strength
mentioned above. $\sigma_{xy}$ for arbitrary Fermi surface is shown
explicitly in Fig.\ref{fermi}(b).

However the suppression of the Hall conductance is not the only
penalty to pay. Right now the Fermi surface is intersecting the edge
state, so the surface is metallic instead of the ideal insulating
one mentioned above. In this case, the electric field in plane would
be greatly reduced by the screening effect, leading to an additional
suppression of the magneto-electric field. Due to the low and
slowly-varying nature of the potential, Thomas-Fermi approximation
is employed. Assume the charge density $\rho({\bf r})$ on the TI
surface is given by
\begin{equation}
\rho({\mathbf r})=-N_f\phi({\bf r})
\end{equation}
where ${\mathbf r}$ is the 2D vector in plane, $\phi({\mathbf r})$
is the scalar potential, and $N_f=e^2E/[2\pi(v_F\hbar)^2]$ is the
density of states. Here $v_F$ is the Fermi velocity of the Dirac
fermion. By the method of Green's function, we can derive the
self-consistent equation in the presence of a point charge $q$ with
distance $R$ away from the TI surface:
\begin{equation}
\varepsilon_0\phi(\mathbf{r},z)=\int d^2\mathbf{r}^{\prime}\frac{\rho(\mathbf{r}%
^{\prime})}{4\pi\sqrt{z^2+(\mathbf{r}-\mathbf{r}^{\prime})^2}}+\frac{q}{4\pi}%
\frac{1}{\sqrt{(z-R)^2+\mathbf{r}^2}}
\end{equation}
Taking the limit $z\rightarrow0$, and applying the Fourier
transformation, we finally get the polar symmetric potential
distribution, given by
\begin{equation}\label{potential}
\phi(r)=\frac{q}{\varepsilon_0}\int_0^\infty \frac{d
k}{(2\pi)^2}\frac{k\exp(-k R)}{2k+1/\ell}J_0(k r)
\end{equation}
where $J_0(x)$ is the zero-order Bessel function, and
$\ell=\varepsilon/N_f$ is the screening length. Then the electric
field in plane can be derived as
$E(\mathbf{r})=-\nabla\phi(\mathbf{r})$. As long as the electric
field in plane is derived, the calculation of TME is
straightforward. The monopole picture recovers when
$\ell\rightarrow\infty$. It's worth emphasizing that although TME
effect survives for finite $\ell$, the monopole picture should be
replaced by magnetic dipole's picture then, and it's explicitly
shown in Eq.(\ref{potential}) that the total effect is further
reduced.


\begin{figure}[tbp]
\begin{center}
\scalebox{0.8}{
\includegraphics[width=10cm,clip]{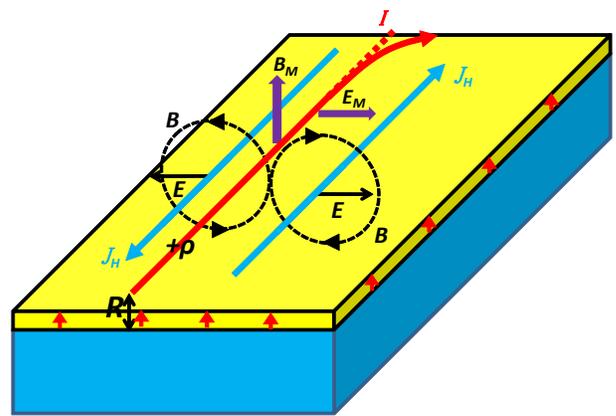}
}
\end{center}
\caption{(Color online) Schematic illustrations of the transverse
force acting on the current. A charged current $I$ (red line) with
charge density $\rho$ is introduced above the gapped TI surface. Its
in-plane electric field $E$ generates Hall current $J_H$ (blue line)
in parallel with the original current. Consequently, an orbital
magnetization $B_M$ (labeled by purple) is induced upward. The
Lorentz force related to $B_M$ induces an unconventional Hall
response $E_M$ transversely. } \label{Efield}
\end{figure}


As a result, this highly nontrivial TME induced by TI is
unfortunately not only governed by the small number
$\alpha\approx1/137$, but also further reduced by several factors
when the realistic situation is considered. So that TME is quite
difficult to be observed experimentally. A way out of this
embarrassing situation is to replace the original point charge by a
charged current flowing in parallel with the TI surface. It appears
in the following that this modification is not only quantitative,
but also qualitative. Here we have to emphasize that the current and
the charge are different quantities as the usual currents are
neutral. Anyway, we still have certain methods to make the
conducting region charged. While such charged current is available
above the gapped TI surface, an in-plane electric field
perpendicular to the current would be generated, see
Fig.\ref{Efield}. Consequently, orbital magnetization is induced by
the Hall current $J_H$ related to this field. Concerning the
symmetry, this magnetization must be perpendicular to both the
current and TI surface. Starting from Eq.(\ref{potential}), simple
calculation shows this magnetic field is
\begin{equation}\label{magnetizationi}
B_M=\frac{\mu_0\rho\sigma _{xy}}{8\pi\varepsilon_0R}[1-(R/\ell)\exp
(R/\ell)\Gamma (R/\ell)]
\end{equation}
where $\Gamma(x)$ is the Gamma function, $\rho$ is the charge
density, and $R$ is the distance between current and TI surface.
This magnetic field would naturally acts a Lorentz force on the
original current, leading to a Hall response and the deflection of
this charged current transversely. Quantitatively, this Lorentz
force is effectively equivalent to a transverse electric field
$E_M$:
\begin{equation}  \label{electric_field}
E_M=\frac{\mu_0I\sigma _{xy}}{8\pi\varepsilon_0R}[1-(R/\ell)\exp
(R/\ell)\Gamma (R/\ell)]
\end{equation}
This equation is the main result of this paper. The Hall response
induced by this transverse electric field is a unique property of
TI. In the limit $\ell\gg R$ , the leading order gives
$E_M=\frac{\mu_0\sigma _{xy}I}{8\pi\varepsilon_0R}$. So the
monopole's picture is recovered. While $\ell\ll R$ is small,
$E_M=\frac{\mu_0\sigma _{xy}I}{8\pi\varepsilon R}\frac{\ell}{R}$,
and dipole picture instead of monopole picture applies $E_M$ is
proportional to $1/R^2$ here. If $R$ is fixed, this result shows
another reduction factor of $\ell/R$ is required. Assume
$\varepsilon_F=10\mathrm{meV}$, rough estimation gives
$\ell\approx500\mathrm{nm}$ for Bi$_2$Se$_3$\cite{Xue2009a}. As a
result, small $\ell\ll R$ limit is adopted usually.

The previous arguments are applied in the laboratory frame. The
physics behind this phenomenon can be even better understood if we
check what's going on in the frame comoving with the charged
current. For simplicity, assume the Fermi surface is in the magnetic
gap. In the comoving frame, the charges are static while the TI
surface as a whole is moving backward. By TME, magnetic monopoles
exist in the mirror. However, as shown before, the physics behind
these monopoles are the quantum Hall effect on the TI surface. As a
consequence, these monopoles are attached to the TI surface, and are
moving backward as well. Motion of the monopoles constructs a
monopole current $I_M$ in the comoving frame. To some extent, the
electric-magnetic duality of Maxwell theory is completely recovered
here. In analogy with the Ampere's law, this monopole current
generates an electric field winding around, which provides a
horizontal but transverse electric field acting on the original
current. This field is exactly the effective field
(Eq.(\ref{electric_field})) derived above.


In the realistic situation, the width $d$ of the conducting region
should be considered. Generically, $d$ is larger than the
current-surface separation, as well as the screening length $\ell$.
Concerning this, detailed calculation shows an additional factor of
$R/d$ should be included in Eq.(\ref{electric_field}). So in the
small $R$ limit, the anomalous electric field approaches to an
constant value proportional to the current density. And in the limit
$\ell\ll R\ll d$, the anomalous field will be proportional to $1/R$.
This result can be understood as follows. The quantity $I/R$ in the
leading term of l.h.s of Eq.(\ref{electric_field}) gives the
dimension of current density. When the large width limit is
considered, this quantity should be replaced by the planer current
density $I/d$. Therefore, a factor of $R/d$ is required. Except for
a quantity close to the unity, we may actually replace $I/R$ by
$I/Max(R,d)$ for simplicity.

Experimentally, the required charged current can be provided by the
steady electron beam emitting from low-energy electron gun (LEED for
example). While drifting above the TI surface, the induced anomalous
electric field would significantly deflect the trajectory of the
electron beam. Numerical estimation shows when the sample size is
$1\mathrm{cm}\times1\mathrm{cm}$, electron velocity is
$1\times10^{5}\mathrm{m/s}$, $m=1\mathrm{meV}$, $I=1\mu\mathrm{A}$,
$d=R=1\mu\mathrm{m}$, the resulting transverse drift would be $5\mu
\mathrm{m}$. These values are realistic ones for beams produced by
electron guns. This deflection can be easily traced by angle
resolved measurement.

\begin{figure}[tbp]
\begin{center}
\scalebox{0.8}{
\includegraphics[width=10cm,clip]{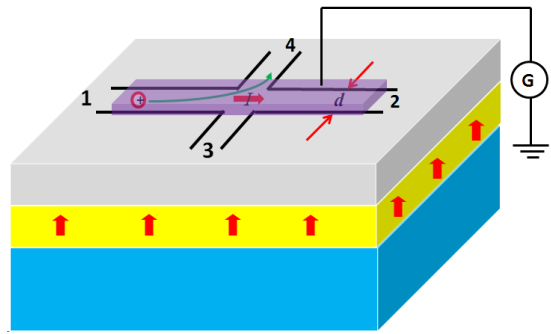}
}
\end{center}
\caption{ Schematic illustrations of UAHE. The bottom blue layer is
TI, the middle one is a magnetic layer, followed by a semiconductor
layer. On the top a gate electrode is deposited. The current is
flowing rightwards, and Hall voltage can be detected transversely. }
\label{ahe}
\end{figure}


In fact, the force given by the monopole current is not the only
force acting on the original current. The original current generates
an image current as well, which would support a Lorentz force acting
on the original current itself. However this Lorentz force is
actually pointing vertically, which is orthogonal to our anomalous
force. As a result, these two effects can be distinguished easily.

While the original current is provided by a quantum wire deposited
on the TI surface, a Hall-like effect can be observed, shown in
Fig.\ref{ahe}. The first layer on the top of TI surface is a
magnetic layer with the magnetization pointing vertically. The
second layer is a semiconductor layer on which quantum wire is
deposited. Employ the usual four-terminal measurement, where
electrodes 1 and 2 are source and drain respectively. Electric field
is measured along electrodes 3 and 4. Gate electrode is deposited on
the top the set up. When it is gated, we may succeed realizing a
charged conducting region in the quantum wire. In equilibrium, this
measured electric field is just the transverse electric field
derived in Eq.(\ref{electric_field}). Usually it is the Hall
resistance $R_H$ that is directly measured. Taking account all the
factors concerned before, in the realistic case $\ell\ll R\ll d$, we
have
\begin{equation}
R_H=\frac{h}{e^2}\alpha^2\frac{m}{\varepsilon_F}\frac{\ell}{R}
\end{equation}
This result shows that the Hall resistance decays inversely
proportional to $R$, which serves as a quantitative characteristic
for this monopole-induced Hall effect. If $R=1\mu\mathrm{m}$, and
the longitude current $I=1\mathrm{mA}$, we have the rough estimation
that $R_H\approx0.01\Omega$, and $V_H\approx10\mu\mathrm{V}$. This
Hall voltage can be easily measured by usual voltmeters. Actually
these conditions can be further optimized. The most effective way is
to drive Fermi surface closer to the Dirac point, as the Hall
voltage increases with $\propto1/\varepsilon_F^2$. Consequently this
effect is quite promising to be detected in the near future.

Conventional anomalous Hall effect (AHE) in ferromagnetic metals has
been well studied in the past\cite{Nagaosa2009a}. In that case, the
magnetization alone is not sufficient to support the giant Hall
effect. Its role is to break the time reversal only, while it is the
spin-orbit interaction that provides the driving force. The
situation is similar in our setup. The ferromagnetic layer on top of
TI surface only breaks the time reversal symmetry and fixes the
chirality. The real driving force is the transverse electric field
given by the monopole current. In this sense we name our new effect
as {\it unconventional anomalous Hall effect}(UAHE). Actually the
magnetic field provided by the magnetic layer is vanishingly small
in the Hall bar measured, and the conventional Hall effect is
negligible here. This magnetic field is restricted on the TI surface
only.

In addition, UAHE is fundamentally different from the conventional
Hall effect. For the conventional one, the Hall voltage satisfies
$V_H=vBd\propto I/\rho$, where $\rho$ is the charge density. While
in the present situation, $V_H\propto \rho v\propto I$. It means
that the Hall voltage here is unchanged as long as the current is
fixed. On the contrary, in the conventional Hall effect, the charge
density matters. If the gate changes sign, charge density and
consequently Hall voltage acquire a sign change as well. From this
point of view, we can easily distinguish UAHE from conventional Hall
effect. Actually, even when the chirality is fixed by an external
magnetic field, where UAHE coexists with conventional Hall effect,
one can also separate UAHE from the conventional one effectively. In
this case, we may adjust the gate voltage to vary the carrier
density, and plot the Hall voltage versus inverse of the carrier
density. The intercept gives the UAHE. On the other hand, we may
also vary the magnetic field. The zero field limit gives the desired
result as well.

It's a revealing issue when the ferromagnetic layer on TI surface is
metallic, and Hall measurement is applied to this layer directly. In
this case, one can have a system with coexistence of AHE and UAHE.
When the layer is thin, the effective current-surface separation is
small, so that anomalous electric field is large. Meanwhile, the
phonon scattering greatly suppresses AHE. As a result, UAHE is
overwhelming, and the net Hall conductance decreases if the layer
thinkness increases. However, in the thick layer limit, UAHE is
vanishingly small, and AHE is dominant. The Hall conductance would
approach to a constant value. This cross-over between AHE and UAHE
help us to distinguish these two effect not only conceptually, but
also experimentally.

In conclusion, we have proposed an unconventional anomalous Hall
effect in this work. In the laboratory frame, the upward magnetic
field induced by a charged current leads to an unconventional Hall
response. This effect can be explained as a monopole current in the
comoving frame. This UAHE survives even when the chemical potential
is away from the gap opened by chirality fixing ferromagnetic layer
on top of TI. Two experiments are proposed in this paper, which
hopefully provide the smoking-guns of TI.

It's also an interesting issue when the chirality of TI surface is
fixed by the spin of the charge carrier itself. When the
current-surface separation is quite small, the local magnetic field
is provided by the local spin only. Opposite spins would lead to
opposite chiralities, and the direction of local monopole current is
therefore opposite. As the charge here is the same for both spin,
the transverse force would be opposite. Consequently, the
conventional spin Hall effect emerges and the transverse spin
voltage is expected. The issue can be simplified when the incident
current is spin polarized, where UAHE induced spin Hall effect would
lead to a Hall voltage built between the two edges. However it
should be pointed out that the gap opened by a single spin is pretty
small ($T(\mathrm{K})\propto R(\mathrm{m})^{-3}$, and $T\sim 1$K
when $R=1{\AA}$), and ultra-low temperature is called for.

We thank M. Kawasaki, P. A. Lee, Y. Tokura and D. Vanderbilt for
insightful discussions. This work is supported by Grant-in-Aids from
under the Grant No.\ 21244053 No. \ 17105002, No. 19048015 and No.
19048008 from MEXT, Japan and Grand in  Ministry of Education of
China under the Grant No. B06011.

\end{document}